\documentclass[twocolumn,twoside,slac_two]{revtex4}
\usepackage{graphicx}
\usepackage{fancyhdr}
\usepackage{hyperref}
\hypersetup{
    colorlinks=true,
    urlcolor=magenta,
}

\begin{document}

\title{SNRs W28 and W44: old cosmic ray accelerators in molecular clouds }

%

\author{V.N.Zirakashvili, V.S.Ptuskin}
\affiliation{Pushkov Institute of Terrestrial Magnetism, Ionosphere
and Radiowave Propagation, 108840 Moscow Troitsk, Russia}

\begin{abstract}
Nonlinear model of diffusive shock acceleration is used for
investigation of the particle acceleration in old supernova remnants
W28 and W44. We modeled the hydrodynamical evolution of the
remnants, shock modification and streaming instability produced by
accelerated particles. Comparison with available radio and gamma-ray
data is given.
\end{abstract}

\maketitle

\thispagestyle{fancy}


\section{Introduction}

The diffusive shock acceleration (DSA) process \cite{krymsky77,bell78,axford77,blandford78} is
considered as the principal mechanism for production of
galactic cosmic rays (CR) in supernova remnants (SNRs). During the last decade
the excellent results of X-ray and gamma-ray astronomy supplied
the observational evidence of the presence of multi-TeV energetic
particles in these objects (see e.g. \cite{lemoine14}).

Most of existing DSA models were applied to young SNRs (see however \cite{lee15}). This is probably
 because it is expected that CRs with  highest energies are produced there.
However lower energy particles are produced  in old SNRs either. So the investigation
 of CR acceleration in old SNRs is important for calculation of overall CR spectra produced
 by SNRs.

In this paper we describe the modifications of our non-linear
 DSA model \cite{zirakashvili12} designed for investigation of DSA in old SNRs. We apply
 it for two GeV bright old SNRs W28 and W44.
%

\section{Nonlinear diffusive shock acceleration model}

Details of our model of nonlinear DSA can
be found in \cite{zirakashvili12}. The model contains coupled
spherically symmetric hydrodynamic equations  and the transport
equations for energetic protons, ions and electrons. The forward
and reverse shocks are included in the consideration.

Damping of magnetohydrodynamic (MHD) waves due to the presence of neutral atoms is important for old SNRs. To take
 this effect into account we add the equation that describes the transport and generation of MHD waves
(see Eq.(4) below).

The hydrodynamical equations for the gas density  $\rho (r,t)$, gas velocity $u(r,t)$,
gas pressure
$P_g(r,t)$, wave pressure $P_m(r,t)$, and the equation for isotropic part of the cosmic ray proton momentum
distribution
 $N(r,t,p)$ in the spherically symmetrical  case are given by

\begin{equation}
\frac {\partial \rho }{\partial t}=-\frac {1}{r^2}\frac {\partial }{\partial r}r^2u\rho
\end{equation}

\begin{equation}
\frac {\partial u}{\partial t}=-u\frac {\partial u}{\partial r}-\frac {1}{\rho }
\left( \frac {\partial P_g}{\partial r}+\frac {\partial P_c}{\partial r}
+\frac {\partial P_m}{\partial r}\right)
\end{equation}

\begin{eqnarray}
\frac 1{\gamma _g-1}\left( \frac {\partial P_g}{\partial t}+u\frac {\partial P_g}{\partial r}
+\frac {\gamma _gP_g}{r^2}\frac {\partial r^2u}{\partial r}\right) = \nonumber \\
-\Lambda (T_e)n^2+H_{c}+2\Gamma _n\frac {P_m}{\gamma _m-1}-\xi _AV_{Ar}(1-h_m)\frac {\partial P_c}{\partial r}
\end{eqnarray}

\begin{eqnarray}
\frac {\partial P_m}{\partial t}+(u+\xi _AV_{Ar})\frac {\partial P_m}{\partial r}
+\frac {P_m}{r^2}\frac {\partial r^2(\gamma _mu+\xi _AV_{Ar})}{\partial r}=  \nonumber \\
-h_m(\gamma _m-1)\xi _AV_{Ar}\frac {\partial P_c}{\partial r}-2\Gamma _nP_m
\end{eqnarray}

\begin{eqnarray}
\frac {\partial N}{\partial t}=\frac {1}{r^2}\frac {\partial }{\partial r}r^2D(p,r,t)
\frac {\partial N}{\partial r}
-w\frac {\partial N}{\partial r}+\frac {\partial N}{\partial p}
\frac {p}{3r^2}\frac {\partial r^2w}{\partial r} \nonumber \\
+\frac 1{p^2}\frac {\partial }{\partial p}p^2b(p)N+\nonumber \\
\frac {\eta _f\delta (p-p_{f})}{4\pi p^2_{f}m}\rho (R_f+0,t)(\dot{R}_f-u(R_f+0,t))\delta (r-R_f(t)) \nonumber \\
+\frac {\eta _b\delta (p-p_{b})}{4\pi p^2_{b}m}\rho (R_b-0,t)(u(R_b-0,t)-\dot{R}_b)\delta (r-R_b(t))
\end{eqnarray}
Here $P_c=4\pi \int dpp^3vN/3$ is the cosmic ray pressure,
$w(r,t)$ is the advection velocity of cosmic rays, $T_e$, $\gamma
_g$ and $n$ are the gas temperature, adiabatic index and number
density respectively,  $\gamma _m$ is the wave adiabatic index, $D(r,t,p)$ is the cosmic ray diffusion
coefficient. The radiative cooling of gas is described by the
cooling function $\Lambda (T_e)$.
 The function $b(p)$ describes the energy losses of particles.
In particular the Coulomb  losses of sub GeV ions
  and the radiative cooling are important in old SNRs. The energy of
 sub GeV ions goes to the gas heating described by the term $H_c$ in Eq. (3).

Cosmic ray diffusion is determined by particle scattering on magnetic inhomogeneities.
The cosmic ray
streaming instability increases the
level of MHD turbulence in the shock
vicinity \cite{bell78} and even significantly amplifies the
absolute value of magnetic field in young SNRs \cite{bell04,zirakashvili08}.
It
decreases the diffusion coefficient and increases the maximum energy of
accelerated particles.
The results of continuing theoretical study of this effect can be found in review papers
\cite{bell2014,Caprioli2014}.

Cosmic ray particles are scattered by moving waves and it is why the cosmic ray advection
velocity
 $w$ may differ from the gas velocity $u$ by the value of the radial
component of the Alfv\'en velocity
$V_{Ar}=V_A/\sqrt{3}$ calculated in the isotropic random magnetic field:
$w=u+\xi _AV_{Ar}$. The factor $\xi _A$
describes the possible deviation of the cosmic ray drift velocity from the gas velocity.
We  use values $\xi _A=1$ and $\xi _A=-1$ upstream of the
forward and reverse shocks respectively, where Alfv\'en waves are
generated by the cosmic ray streaming instability and propagate in
the corresponding directions.

The pressure of generated waves $P_m$ determines the scattering and diffusion of
 energetic particles with charge $q$, momentum $p$ and speed $v$

\begin{equation}
D=D_B\frac {B^2}{8\pi P_m}, \ D_B=\frac {cpv}{3qB}, \ B=\sqrt{B_0^2+8\pi P_m}
\end{equation}
where $B$ is the total magnetic field strength, while $B_0$ is the
strength of the mean field. At high wave amplitudes the diffusion
coefficient coincides with the  Bohm diffusion coefficient $D_B$.

The parameter $h_m$ in Eqs. (3,4)  describes the fraction of magnetic energy
  produced by streaming instability. We use the following dependence $h_m(B)$

\begin{equation}
h_m=1, \ \frac B{B_0}<3; \ h_m=0.5, \ \frac B{B_0}>3.
\end{equation}
At high amplitudes the waves are damped and the fraction $1-h_m$ of energy goes into
 the gas heating
upstream of the shocks \cite{mckenzie82} that is
described by the last term in Eq. (3).
The heating and wave generation limits the total compression ratio of cosmic ray modified shocks.
In the downstream region of
the forward and reverse shock at $R_b<r<R_f$ we put $\xi _A=0$ and therefore $w=u$.

 In the shock transition region the magnetic pressure
 is increased by a  factor of $\sigma ^{\gamma _m}$,  where
$\sigma $ is the shock compression ratio. Its impact  on the shock
dynamics is taken into account via the Hugoniot conditions.

Below we use the adiabatic index of Alfv\'en waves $\gamma _m=3/2$.
For this value of the adiabatic index, the wave pressure
$P_m=(\delta B)^2/8\pi $ equals to
 the  wave magnetic energy density.

The rate of the neutral damping $\Gamma _n=0.5\sigma
_{ex}\mathrm{v}_Tn_n$ depends on the
 number density of neutrals $n_n$, thermal speed $\mathrm{v}_T\sim 10$ km s$^{-1}$  and charge exchange
cross-section $\sigma _{ex}\sim 10^{-14}$ cm$^{2}$. We take the damping into account
in the upstream region only and put $\Gamma _n=0$ downstream of the shock.

Two last terms in Eq. (5)
correspond to the injection of thermal protons with momenta
$p=p_{f}$, $p=p_{b}$ and mass $m$ at the forward and
reverse shocks located at $r=R_f(t)$ and $r=R_b(t)$
respectively
. The
dimensionless parameters $\eta _f$ and $\eta _b$ determine the efficiency of injection.

The injection efficiency is taken to be
independent of time $\eta _f=0.001$, and the particle injection
momentum is  $p_{f}=2m(\dot{R}_f-u(R_f+0,t))$.
Protons of mass $m$ are
injected at the forward shock.
The high injection efficiency
results in the significant shock modification already at early stage of
SNR expansion. Since in the old remnants the reverse shock absent we
  put $\eta _b=0$ in the modeling below.

We neglect the pressure of energetic electrons and treat them as test particles.
The evolution of the
electron distribution is described by equation analogous to Eq. (5)
with function $b(p)$  describing Coulomb, synchrotron and inverse Compton
(IC) losses and additional terms describing the production of secondary
leptons by energetic protons and nuclei.

\section{Modeling of diffusive shock acceleration in the old SNRs}

A significant part of core collapse supernova explosion occurs in molecular gas. The stars with
 initial masses below $12\ M_{\odot}$ have no power stellar winds and therefore do not produce a strong
 modification of their circumstellar medium. The molecular cloud have been totally destroyed by stellar winds and supernova
 explosions  of more massive stars at the instant of explosion. As a result the star explodes in the inter-clump
 medium with the density $5-25$ cm$^{-3}$ \cite{chevalier99}.  Many such SNRs are observed in gamma rays now.

\begin{table}[t]
\begin{center}
\caption{Physical parameters of SNRs W28 and W44}
\begin{tabular}{|c|c|c|c|c|c|c|c|c|c|c|}
\hline     & $d$ & $R_f$  & $E_{SN}$ & $M_{ej}$ &  $n_H$ &  $n_n$  & $K_{ep}$ & $T$  & $V_f$ & $B_f$\\
\hline     &kpc& pc & $10^{51}$erg &$M_{\odot }$   &cm$^{-3}$ &cm$^{-3}$& &kyr&km/s&$\mu $G\\
\hline W28 &1.9& 13.4 & 1.3    & 6.8   & 4.0  & 0.2   & 0.008   & 37 &121  &79  \\
\hline W44 &2.8& 12.4 & 1.6    & 7.1   & 6.0  & 0.3   & 0.006   & 33 &130  &102   \\
\hline
\end{tabular}
\label{tab:k}
\end{center}
\end{table}

Bright in GeV gamma rays SNRs W44 and W28 at distance $d\sim 2-3$ kpc from the Earth are at the radiative phase now
and show signs of interaction with molecular gas \cite{reach05}. It is believed that circumstellar medium  is almost fully
 ionized by ultraviolet radiation from the remnant interior at the radiative stage \cite{chevalier99}.
The same is true for young SNRs
 because the gas is ionized by the radiation from the shock breakout and in the hot shock precursor produced
by accelerated particles. Probably there exist an intermediate phase
when the shock propagates in the neutral medium. We leave the detail
description of the gas ionization to the future work. Here we shall
use a simplified approach and
 consider the number density of neutrals $n_n$ as a free parameter.

\begin{figure}
\begin{center}
\includegraphics[width=8.0cm]{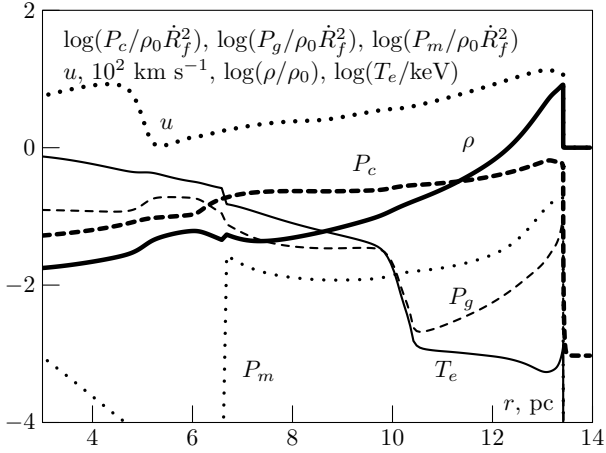}
\end{center}
\caption{Radial dependencies of the gas density (thick solid line), the gas
velocity (dotted line), CR pressure (thick dashed line),
the wave magnetic energy density $(\delta B)^2/8\pi =P_m$ (thin dotted line),
the gas
temperature $T_e$ (thin solid line)
and the gas pressure $P_g$ (dashed line) at
 $T=37$ kyr in SNR W28.}
\end{figure}

The parameters of supernova modeling are given in Table I. The
explosion energy $E_{SN}$ was taken within the range $(1-2)\cdot
10^{51}$ erg. The ambient number density $n_H$ was adjusted to
reproduce the observable gamma ray
 fluxes.
The number density of neutrals was adjusted to reproduce the
spectral shape of gamma emission.
The remnant evolution was calculated up to the instant of time $T$ when the radius of the forward shock equals to the
observable radius $R_f$.
The electron to proton ratio $K_{ep}$ was adjusted to
reproduce the observable radio-flux. The numbers in three last
columns of Table I that is the age $T$, shock speed $V_f$ and magnetic field strength $B_f$  just downstream of the shock
were obtained in the modeling.



Figures (1)-(5) illustrate the results of our numerical calculations.

\begin{figure}
\includegraphics[width=8.0cm]{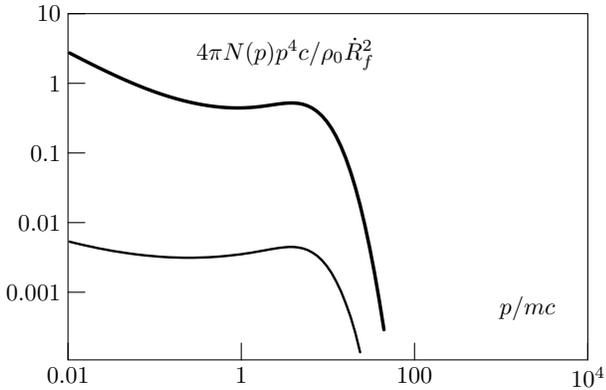}
\caption{Spectra of accelerated particles at the forward shock in SNR W28 at $T=37$ kyr.  The
spectrum of protons (thick solid line) and electrons (thin solid line) are shown. }
\end{figure}


Radial dependencies of physical quantities in SNR W28 at present
 ($T=37$ kyr) are shown in Fig.1. The contact discontinuity between the ejecta and
the interstellar gas is at $r=R_c=6.6$ pc. The gas temperature drops
sharply downstream of the forward shock due to the radiative
cooling. However a thin dense shell is not formed
 because of CR pressure (cf. \cite{lee15}). The central part of the remnant is filled
by the hot rarefied gas with temperature $10^6-10^7$K.

\begin{figure}
\includegraphics[width=8.0cm]{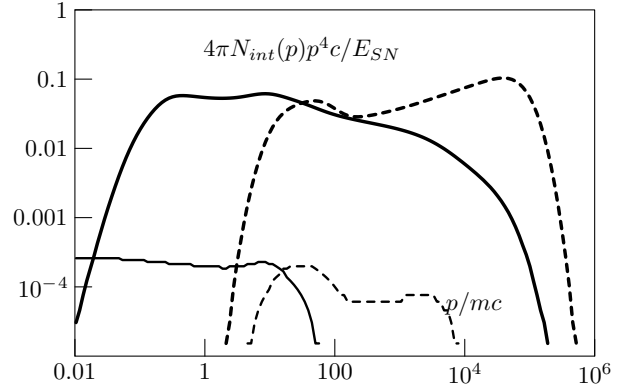}
\caption{ Spectra of particles produced in the SNR W28
during $37$ kyr after explosion. The spatially integrated spectrum
of protons (thick solid line), the spectrum of protons
escaped from the remnant (dashed line), the integrated spectrum of
electrons
 (thin solid line) and the spectrum of electrons
escaped from the remnant (thin dashed line) are shown.}
\end{figure}

\begin{figure}
\includegraphics[width=8.0cm]{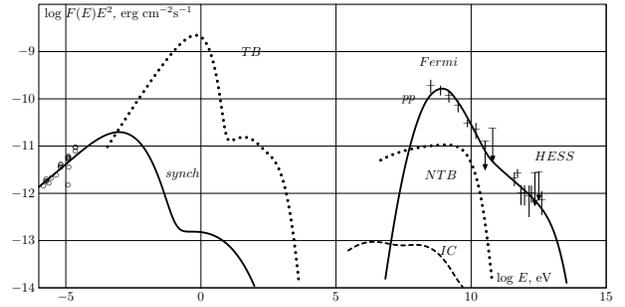}
\caption{The results of modeling of electromagnetic radiation of W28.
The following
radiation processes are taken into account: synchrotron radiation
of accelerated electrons (solid curve on the left), IC emission
(dashed line), gamma-ray emission from pion decay (solid line on
the right), thermal bremsstrahlung (dotted line on the left), nonthermal bremsstrahlung
(dotted line on the right).
 Experimental
data in  gamma-ray Fermi LAT \cite{abdo10}; HESS \cite{aharonian08} (data
with error-bars) and
radio-bands \cite{dubner00} (circles)  are also shown. }
\end{figure}

Spectra of accelerated in W28 protons and electrons at $T=37$ kyr
are shown in Fig.2. At this point the maximum energy of accelerated
protons is about $10$ GeV. This value  is in accordance with with
the estimate of the maximum energy
 in the partially ionized medium $E_{\max }=u^3_8n_H^{1/2}n_n^{-1}$ TeV \cite{drury96}.
 For shock velocity $V_f\sim 100$ km s$^{-1}$ that is $u_8=0.1$ and using the parameters from the Table I we indeed get
$E_{\max }\sim 10$ GeV.

The spectra of particles $N_{int}$ produced during  37 kyr after supernova explosion
 are shown in Fig.3. They are calculated via the integration throughout the simulation domain and via the
integration on time of  the outward diffusive flux at the simulation
boundary at $r=2R_f$. About $85\%$ of the kinetic energy of explosion is transferred
to cosmic rays. Most of this energy is gone by escaped particles.
The maximum energy of escaped particles is 100 TeV for this SNR.

\begin{figure}
\includegraphics[width=8.0cm]{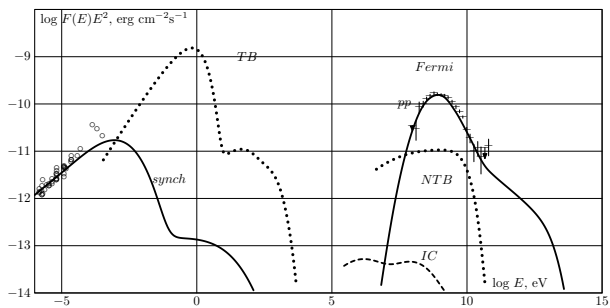}
\caption{The results of modeling of electromagnetic radiation of W44.
The following
radiation processes are taken into account: synchrotron radiation
of accelerated electrons (solid curve on the left), IC emission
(dashed line), gamma-ray emission from pion decay (solid line on
the right), thermal bremsstrahlung (dotted line on the left), nonthermal bremsstrahlung
(dotted line on the right).
 Experimental
data in  gamma-ray  Fermi LAT \cite{ackermann13}, (data
with error-bars) and
radio-bands \cite{castelletti07,arnaud16}  (circles) are also shown. }
\end{figure}

While the maximum energy of protons at the shock is about 10 GeV the
interior of SNR contains protons with energies up to several TeV. They
were accelerated earlier when the shock speed was higher.

Results of multi-band modeling of SNRs W28 and W44 are shown in
Figures 4,5. Thermal emission shows two components. One is produced
by the hot gas in the remnant interior while lower energy component
is produced by the dense gas radiatively cooled and recombined
behind the shock front.

\section{Discussion}
High acceleration efficiency ($85\%$) in our modeling  seems at
odds with the lower energetics of $10-20\%$ expected for Galactic SNRs.
Higher density of circumstellar medium might permit the
lower acceleration efficiency. However this will result in the
over-production of the thermal radio emission in SNR W44. Indeed the
thermal radio-flux is only slightly below than the {\it Planck} data
\cite{arnaud16} at high radio-frequencies (see Fig. 5). In this regard
 the direct DSA scenario seems more probable in comparison with the reacceleration
 scenario in SNR W44 \cite{cardillo16}.

The main part of gamma emission in W28 and W44 is produced by energetic protons via $pp$ collisions.
The gamma emission at GeV energies is
 produced by protons recently accelerated at the forward shock. Their energy is regulated by the neutral damping of
MHD waves upstream of the shock.
Higher energy gamma emission is produced by particles accelerated
 earlier when the shock speed was higher. The confinement of these particles is very efficient because of the Bohm like diffusion
in the ionized  remnant interior.

So we conclude that the neutral damping naturally results in appearance
of high energy tails of gamma emission in old SNRs.

The work was supported by Russian Foundation of Fundamental Research grant 16-02-00255.

\bigskip 

\begin{thebibliography}{99}   


\bibitem{krymsky77} Krymsky, G.F. 1977, Soviet Physics-Doklady, 22, 327

\bibitem{bell78} Bell, A.R., 1978, MNRAS, 182, 147

\bibitem{axford77} Axford, W.I., Leer, E. \& Skadron, G., 1977, Proc. 15th
ICRC, Plovdiv, 90, 937

\bibitem{blandford78} Blandford, R.D., \& Ostriker, J.P. 1978, ApJ, 221, L29

\bibitem{lemoine14} Lemoine-Goumard M. Proceedings of  IAU Symposium 2014, 296, 287

\bibitem{lee15} Lee S.H., Patnaude D.J.,  Raymond J.C. et al. 2015, ApJ 806, 71


\bibitem{zirakashvili12} Zirakashvili, V.N. \& Ptuskin V.S. 2012, Astropart. Phys., 39, 12




\bibitem{bell04} Bell, A.R., 2004, MNRAS, 353, 550

\bibitem{zirakashvili08} Zirakashvili, V.N., \& Ptuskin, V.S., 2008, ApJ, 678, 939

\bibitem{bell2014} Bell, A.R., 2014, Astropart. Phys., 43, 56

\bibitem{Caprioli2014} Caprioli, D., 2014, Nuclear Physics B (Proc. Suppl.), 256, 48

\bibitem{mckenzie82} McKenzie, J.F., \& V\"olk, H.J., 1982, A\&A, 116, 191


\bibitem{chevalier99} Chevalier R. 1999, Astrophys. J. 511, 798.

\bibitem{reach05} Reach, W.T., Rho, J., \& Jarrett, T.H., 2005, ApJ 618,297





\bibitem{drury96} Drury, L.O., Duffy, P., \& Kirk, J. G. 1996, A\&A, 309, 1002








\bibitem{abdo10} Abdo, A.A., Ackermann, M., Ajello, M., et al. 2010, ApJ, 718, 348

\bibitem{aharonian08} Aharonian, F., et al. 2008, A\&A, 481, 401

\bibitem{dubner00} Dubner, G.M., Velazquez, P.F., Goss, W.M., \& Holdaway, M.A., 2000, ApJ, 120, 1933

\bibitem{ackermann13} Ackermann, M., Ajello, M., Allafort, A., et al. 2013, Sci, 339, 807

\bibitem{castelletti07} Castelletti1, G., Dubner, G., Brogan, C., \& Kassim, N.E., 2007, A\&A 471, 537

\bibitem{arnaud16} Arnaud M., Ashdown M., Atrio-Barandela F. et al., 2016, A\&A 586, 134

\bibitem{cardillo16} Cardillo, M., Amato, E., \& Blasi, P., 2016, A\&A 595, 58






\end{thebibliography}

\end{document}